\input vanilla.sty
\magnification 1200
\centerline{\bf p-ADIC AND ADELIC GENERALIZATION}
\centerline{\bf OF QUANTUM COSMOLOGY}
\vskip 8pt

\centerline{Branko Dragovich$^*$,}
\centerline{Ljubi\v sa Ne\v si\'c$^{**}$}
\vskip 8pt
\centerline{\it $^*$Institute of Physics, P.O.Box 57, 11001 Belgrade,
Yugoslavia,}

\centerline{\it Steklov Mathematical Institute, Moscow, Russia}
\centerline{\it E-mail: dragovic\@phy.bg.ac.yu}
       
\centerline{\it $^{**}$Department of Physics, University of Ni\v s, P.O. Box 91,
18000 Ni\v s, Yugoslavia}
\centerline{\it E-mail: nesiclj\@junis.ni.ac.yu}

\vskip 12pt
\noindent
\bf 1. Abstract
\vskip 12pt

\rm $p$-Adic and adelic generalization of ordinary
quantum cosmology is considered.
In [1], we have calculated $p$-adic  wave functions for 
some minisuperspace cosmological models according to the 
"no-boundary" Hartle-Hawking proposal.
In this article, applying $p$-adic and adelic quantum mechanics, 
we show existence of the corresponding vacuum eigenstates.
Adelic wave function contains some information on discrete structure 
of space-time at the Planck scale.


\vskip 12pt
\noindent
\bf 1. Introduction
\vskip 12pt
\baselineskip 12pt
\rm While measurement data always belong to the field 
of rational numbers $Q$, by mathematical reasons, standard physical models
use the field of real numbers $R\equiv Q_\infty$,
and the field of complex numbers $C$. Completion of $Q$
with respect to the
absolute value ($|\cdot |_\infty$) 
gives $R$, and its algebraic extension yields $C$. According 
to the Ostrowski theorem [2] any non-trivial norm on the field of rational
numbers $Q$ is equivalent to the usual absolute value $|\cdot |_\infty$ 
or 
to the $p$-adic norm $|\cdot |_ p$, $p=$a prime number.
$p$-Adic norm is the non-Archimedean (ultrametric) one and for a 
rational number, $0\ne x\in Q$, 
$x=p^\gamma {m \over n}$,
$0\ne n,\gamma, m\in Z$, has the value $|x|_p=p^{- \gamma}$.
Completing  $Q$ with respect to the $p$-adic norm one gets
the field of $p$-adic numbers $Q_p$.

According to quantum gravity there is an uncertainty of measuring 
distances [2,3]:
$\Delta l\geq l_{pl}=\sqrt{\hbar G/c^3},$
where $l_{pl}\sim 10^{-33}cm$ is the Planck lenght, 
$\hbar=\frac{h}{2\pi}$ is the reduced Planck 
constant, $G$ is Newton's gravitational constant and $c$ is the speed of
light. 
Thus the Planck lenght is the smallest possible distance which can be 
in principle  measured. For these very small distances the Archimedean 
axiom of the Euclidean geometry is no more valid. Impossibility to
measure the Archimedean distances shorter than the Planck lenght 
and possible existence
of non-Archimedean spaces at the Planck scale  is one of the main
motivations for the investigation of $p$-adic quantum models.

After successful application of $p$-adic models to string amplitudes in
1987 [3-5],
and formulation of $p$-adic quantum mechanics [6-8], one-dimensional
systems with quadratic action were considered:
a free particle and a harmonic oscillator [2],
a particle in a constant field [9], a harmonic oscillator with
time-dependent
frequency [10].

In $p$-adic quantum mechanics with complex-valued wave functions, the
Schr\"odinger equation cannot be written down, because
$x\in Q_p$, $\psi(x)\in C$
and derivative $\frac{d\psi}{dx}$ as well as product $x\psi(x)$ have no
sense. 
However, finite transformations
are meaningful and the corresponding Weyl and evolution operators
are $p$-adically well defined. Ordinary and $p$-adic quantum mechanics
are
unified within adelic quantum mechanics [11].

The application of $p$-adic numbers and adeles in quantum cosmology is
of special interest.
It should give a new information about structure of space and time,
as well as more profound approach to the very early universe.

\noindent
\vskip 12pt
\noindent
\bf 2. \it{\bf{p}}\bf{-Adic numbers and adeles}
\vskip 12pt
\rm $p$-Adic number $x\in Q_p$, in the canonical form, is an infinite
expansion
$$
x=p^\gamma\sum\limits_{i=0}\limits^\infty x_ip^i,\qquad x_0\ne0,\quad 
0\leq x_i\leq p-1.\eqno(2.1)
$$ 
The norm of $p$-adic number $x$ in (2.1) is $|x|_p=p^{-\gamma}$ and
satisfies not
only the triangle inequality but also the stronger inequality
$$|x+y|_p\le\max (|x|_p,|y|_p).\eqno{(2.2)}$$
Metric on $Q_p$ is defined by $d_p(x,y)=|x-y|_p$. This metric is a
non-Archimedean one and pair ($Q_p,d_p$) is locally compact,
complete, separable and totally disconnected metric space.

Real and $p$-adic numbers may be unified in the form of the adeles [12].
An adele is an infinite sequence
$$
a=(a_\infty, a_2,...,a_p,...),\eqno(2.3)
$$
where $a_\infty\in Q_\infty$, and $a_p\in Q_p$,
with restriction
that $a_p\in Z_p$ $(Z_p=\{x\in Q_p:
|x|_p\leq1\})$ for all but a finite number of $p$. 

Let $S$ be a finite set of prime numbers and 
${\cal A}(S)=Q_\infty\times\prod\limits_{p\in S} Q_p\times\prod
\limits_{p \notin S} Z_p$. Space of all adeles is then 
 ${\cal A}=\bigcup\limits_S{\cal A}(S)$ and it is a topological ring.
( ${\cal A}$ is a ring with respect to the componentwise addition and
multiplication.) A principal adele is a sequence $(r,r,...,r,...)\in
{\cal
A}$,
where $r\in Q$.

An important function on ${\cal A}$ is the
additive character $\chi(x), \ x\in {\cal A}$, which is continuous and
complex valued function with properties
$$
|\chi_\upsilon(x_\upsilon)|_\infty=1,
\quad\chi_\upsilon(x_\upsilon+y_\upsilon)=
\chi_\upsilon(x_\upsilon)\chi_\upsilon(y_\upsilon),\eqno(2.4)
$$
$\upsilon\in V=\{\upsilon:\upsilon=\infty,2,...,p,...\}$.
The additive character over the group ${\cal A}$ is given by
$$
\chi(x)=\prod_\upsilon\chi_\upsilon(x_\upsilon)=
\exp(-2\pi ix_\infty)\prod_p
\exp(2\pi i\{x_p\}_p),\eqno(2.5)
$$
where $\{x\}_p$ is the fractional part of a $p$-adic number $x$.

An integral of the Gaussian type over $Q_\upsilon$ is
$$
\int\limits_{Q_\upsilon}\chi_\upsilon(ax^2+bx)dx=
\lambda_\upsilon(a)|2a|^{-1/2}_\upsilon
\chi_\upsilon\left(-\frac{b^2}{4a}\right),\quad a\ne 0,\eqno(2.6)
$$
where $\lambda_\upsilon(a):\enskip  Q_{\upsilon}\mapsto C$,
is the number-theoretical function [2] which has the following
properties:
$$
|\lambda_\upsilon(a)|_\infty=1,\quad 
\lambda_{\upsilon}(0) =1,
\lambda_\upsilon(ab^2)=\lambda_\upsilon(a),\eqno(2.7)
$$
$$
\lambda_\upsilon(a)\lambda_\upsilon(b)=
\lambda_\upsilon(a+b)\lambda_\upsilon(a^{-1}+b^{-1}),\eqno(2.8)
$$
for any $\upsilon=\infty,2,...,p,...$, and $a\ne0,b\ne0$.

$p$-Adic Gauss integral over the region of integration $|x|_p\leq
p^{-\nu}$ is
$$
\enskip
\int\limits_{|x|_p\leq p^{-\nu}}\chi_p(\alpha x^2+\beta x)dx=
\cases 
p^{-\nu}\Omega(p^{-\nu}|\beta|_p),
|\alpha|_pp^{-2\nu}\leq1,\\
\lambda_p(\alpha)|2\alpha|^{-1/2}_p
\chi_p\left(-\frac{\beta^2}{4\alpha}\right)
\Omega\left(p^\nu\left|\frac{\beta}{2\alpha}\right|_p\right)\!\!,
\!\!\!\enskip |\alpha|_pp^{-2\nu}>1,  
\endcases
\eqno(2.9)
$$
where $\Omega (u)$ is defined as follows:
$$
\Omega (u) = \cases 
\! 1, \enskip u \leq 1,\\
\! 0, \enskip u >1.
\endcases
\eqno(2.10)
$$

\vskip 12pt
\noindent
\bf 3. Adelic quantum mechanics
\vskip 12pt

\rm Standard quantum mechanics with complex functions of a real
variable,
$\psi\in L_2(Q_\infty)$,
can be generalized to $p$-adic quantum mechanics with complex functions
of a $p$-adic argument, $\psi\in L_2(Q_p)$.
Both quantum mechanics (real and $p$-adic)
can be unified in the form of {\it adelic
quantum mechanics} with complex functions of  adelic variables,
$\psi\in L_2({\cal A})$ [11].

For a system which classical, real and $p$-adic, dynamics can be
represented in the form
$$
z(t)={\cal T}(t,t_0)z(t_0),\quad\quad
z=
\pmatrix
q\cr
k\cr\endpmatrix
\eqno(3.1)
$$
($z=(z_\infty,z_2,\dots,z_p,\dots)$), adelic quantum
mechanics is a triple
$$
(L_2({\cal A}),W(z),U(t)),
\eqno(3.2)
$$
where $z$ is a point in classical adelic phase space, $t$ is an adelic
time, $L_2({\cal A})$ is the
Hilbert space of complex square-integrable functions with
respect to the Haar measure on ${\cal A}$,
$W(z)$ is a unitary representation of
Heisenberg-Weyl group on $L_2({\cal A})$
and $U(t)$ is a unitary 
representation of the evolution operator on $L_2({\cal A})$.

In equation (3.1), $q$ and $k$ are position and 
momentum, respectively. ${\cal T}$ is a
matrix evolution, which satisfies ${\cal T}(t_2,t_1){\cal T}(t_1,t_0)=
{\cal T}(t_2,t_0)$
and $B({\cal T}(t,t_0)z,$ ${\cal T}(t,t_0)z')=B(z,z')$,
where $B(z,z')=-kq'+qk'$ is 
symplectic bilinear form on the adelic phase space.

The operator $W_\upsilon(z)$
acts on the wave function $\psi_\upsilon(x)\in
L_2(Q_\upsilon)$
in the following way:
$$
W_\upsilon(z)\psi_\upsilon(x)=
\chi_\upsilon(\frac{kq}{2}+kx)\psi_\upsilon(x+q).\eqno(3.3)
$$
An evolution operator $U(t)$ is defined by
$$
U(t)\psi(x)=
\int\limits_A{\cal K}_t(x,y)\psi(y)dy=
\prod_\upsilon
\int\limits_{Q_\upsilon}{\cal K}_{t_\upsilon}^{(\upsilon)}
(x_\upsilon,y_\upsilon)\psi^{(\upsilon)}(y_\upsilon)dy_\upsilon
\eqno(3.4)
$$
and it describes dynamics in adelic quantum mechanics. The operator
$U(t)$ and
its kernel ${\cal K}_{t_\upsilon}^{(\upsilon)}
(x_\upsilon,y_\upsilon)$ satisfy product relations:
$$
U(t+t')=U(t)U(t'),
$$
$$
{\cal K}_{t+t'}^{(\upsilon)}(x,y)=
\int\limits_{Q_\upsilon}{\cal K}_t^{(\upsilon)}(x,z)
{\cal K}_{t'}^{(\upsilon)}(z,y)dz.\eqno(3.5)
$$
By analogy with standard quantum mechanics, kernel of adelic 
evolution operator is given by product of Feynman's path integrals
$$
{\cal K}_t^{(\upsilon)}(x,y)
=\int\chi_\upsilon\left(-\smallint\limits_0\limits^tL_\upsilon(q,\dot q,
t)dt
\right){\cal D}q(t)\eqno(3.6)
$$
(we use $h=1$ for the Planck constant), where integration is 
performed over classical real and $p$-adic trajectories with the
boundary conditions $q(0)=y$, $q(t)=x$.

The eigenvalue problem for $U(t)$ reads
$$
U(t)\psi_{\alpha\beta}(x)=\chi(E_\alpha t)\psi_{\alpha\beta}(x),
\eqno(3.7)
$$
where $\psi_{\alpha\beta}(x)$ are adelic wave eigenfunctions,
$E=(E_\infty,E_2,\dots,E_p,\dots)$ is the corresponding adelic energy,
$\alpha=(\alpha_\infty,\alpha_2,\dots,\alpha_p,\dots)$ and 
$\beta=(\beta_\infty,\beta_2,\dots,\beta_p,\dots)$ are indicies for
energy
levels and their degeneration, respectively.

The eigenstates for adelic evolution operator (3.7) are infinite
products
of eigenfunctions from the corresponding real and $p$-adic counterparts
of
a quantum model.
One has to point out that any adelic eigenfunction contains only
finitely
many $p$-adic eigenfunctions which are different from the vacuum state
$\Omega(|x|_p)$ defined by (2.10).

\vskip 12pt
\noindent
\bf 4. Adelic quantum cosmology
\vskip 12pt

\rm Adelic quantum cosmology is an application of adelic quantum 
theory to the universe as a whole.
In the path integral approach to standard quantum cosmology 
starting point is Feynman's
idea that the amplitude to go from one state with intrinsic metric
$h_{ij}$, and matter configuration $\phi$ on an initial hypersurface
$\Sigma$, to another state with metric $h_{ij}^\prime$, and matter
configuration $\phi^\prime$ on a final hypersurface $\Sigma^\prime$, is
given by a functional integral of
$\chi_\infty(-S_\infty[g_{\mu\nu},\Phi])$ over all four-geometries
$g_{\mu\nu}$, and matter configurations $\Phi$, which interpolate
between
the initial and final configurations [13], i.e.
$$
\langle h_{ij}^\prime,\phi^\prime,\Sigma^\prime|
h_{ij},\phi,\Sigma\rangle_\infty=
\int {\cal D}{(g_{\mu\nu})}_\infty {\cal D}(\Phi)_\infty
\chi_\infty(-S_\infty[g_{\mu\nu},\Phi]).\eqno(4.1)
$$
The $S_\infty[g_{\mu\nu},\Phi]$ is the usual Einstein-Hilbert action
$$
S[g_{\mu\nu},\Phi]=
\frac{1}{16\pi G}\left(\quad\int\limits_M d^4x\sqrt{-g}(R-2\Lambda)
+2\int\limits_{\partial M}d^3x\sqrt{h}K\right)
$$
$$
-\frac{1}{2}\int\limits_M d^4x\sqrt{-g}
\lbrack
g^{\mu\nu}\partial_\mu\Phi\partial_\nu\Phi+V(\Phi)\rbrack\eqno(4.2)
$$
for the gravitational field and matter fields $\Phi$. In (4.2), $R$ is 
scalar curvature of four-manifold $M$, $\Lambda$ is cosmological
constant,
$K$ is trace of the extrinsic curvature
$K_{ij}$ at the boundary $\partial M$ of the manifold $M$. 

To perform $p$-adic and adelic generalization we first make $p$-adic
counterpart of the action (4.2) using form-invariance under change
of real to the $p$-adic number fields [2,14]. Then we generalize (4.1)
and
introduce $p$-adic complex-valued cosmological amplitude
$$
\langle h_{ij}^\prime,\phi^\prime,\!\!\Sigma^\prime|
h_{ij},\phi,\Sigma\rangle_p=\!\!\!
\int\!\!{\cal D}{(g_{\mu\nu})}_p{\cal D}(\Phi)_p
\chi_p(-S_p[g_{\mu\nu},\!\!\Phi]).\eqno(4.3)
$$

The space of all 3-metrics and matter field configurations
$(h_{ij}(\vec{x}),\phi(\vec{x}))$ on a 3-surface is called superspace
(this is the configuration space in quantum cosmology).
Superspace is the infinite dimensional one with a finite number of
coordinates
$(h_{ij}(\vec{x}),\phi(\vec{x}))$ at each point $\vec{x}$ of the
3-surface.
In practice the work with the infinite dimensions is not possible.
One useful approximation therefore is to truncate the infinite degrees
of
freedom to a finite number, thereby obtaining some particular
minisuperspace model.
Usually, one restricts the four-metric to be of the form
$
ds^2=-N^2(t)dt^2+h_{ij}dx^idx^j,
$
where $N(t)$ is the laps function.  Three-metric $h_{ij}$ and matter
fields
are restricted in such
a way that they depend on a finite number of functions of $t$,
$q^\alpha(t)$, 
where $\alpha=1,2,...,n$. 
For such minisuperspaces, functional integrals (4.1) and (4.3) are reduced
to  
functional
integration over three-metrics, matter configurations and to one 
usual integral over the laps function.
If one takes boundary condition $q^\alpha(t_2)=q^\alpha_2$, 
$q^\alpha(t_1)=q^\alpha_1$ then integral in (4.1) and (4.3),
in the gauge $\dot N=0$, is a
minisuperspace propagator. In this case it holds
$$
\langle q^\alpha_2|q^\alpha_1\rangle_\upsilon
=\int dN {\cal K}_\upsilon(q^\alpha_2,N|q^\alpha_1,0),\eqno(4.4)
$$
where
$$
{\cal K}_\upsilon(q^\alpha_2,N|q^\alpha_1,0)
=\int {\cal D}q^\alpha\chi_\upsilon(-S_\upsilon[q^\alpha])\eqno(4.5)
$$
is an ordinary quantum-mechanical propagator between fixed $q^\alpha$ in
fixed time $N$.

For one dimensional quantum systems $p$-adic path integral is
investigated in  [15], where it is shown that for quadratic classical
action $S^{cl}_p(q_2,N|q_1,0)$, (4.5) becomes
$$
{\cal K}_p(q_2,N|q_1,0)=
\lambda_p
\left(
-\frac{\partial^2 S^{cl}_p}{2\partial q_2\partial q_1}
\right)
\left|
\frac{\partial^2S^{cl}_p}{\partial q_2\partial q_1}
\right|_p^{1/2}
\chi_p(-S^{cl}_p(q_2,N|q_1,0)).
\eqno(4.6)
$$

\noindent 
If system has $n$ decoupled
degrees of freedom, its $p$-adic kernel is a product
$$
{\cal K}_p(q_2,N|q_1,0)=\prod\limits_{\alpha=1}\limits^n
\lambda_p
\left(
-\frac{\partial^2 S^{cl}_p}{2\partial q^\alpha_2\partial q^\alpha_1}
\right)
\left|
\frac{\partial^2S^{cl}_p}{\partial q^\alpha_2\partial q^\alpha_1}
\right|_p^{1/2}
\chi_p(-S^{cl}_p(q^\alpha_2,N|q^\alpha_1,0)).
\eqno(4.7)
$$

Expressions (4.6) and (4.7) have the same form as in ordinary
quantum mechanics [15].

$p$-Adic and adelic wave functions of the universe may be found 
by means of the equation (3.7). The corresponding adelic 
eigenstates have the form
$$
\Psi (q^\alpha ) = \psi_\infty (q_\infty^\alpha) \prod_{p\in S}
\psi_p(q_p^\alpha) \prod_{p\not\in S} \Omega (|q_p^\alpha|_p) .
\eqno(4.8)
$$

A necessary condition to construct an adelic model is existence of the
$p$-adic (vacuum) state
$\Omega(|q^\alpha|_p)$, which satisfies
$$
\int\limits_{|q^\alpha_1|_p\leq 1}{\cal K}_p(q^\alpha_2,N|q^\alpha_1,0)
dq^\alpha_1=
\Omega(|q^\alpha_2|_p)\eqno(4.9)
$$
for all but a finite number of $p$.

\vskip 12pt
\noindent
\bf 5. Some \it{\bf{p}}\bf{-adic and adelic minisuperspace models}

\vskip 12pt
\noindent
\it 5.1. p-Adic and adelic model with cosmological constant in $D=3$
dimensions 
\vskip 10pt
\rm This model in the real case is considered in the paper [16]. Its
metric is
$$
ds^2=\sigma^2\left(-N^2(t)dt^2+a^2(t)(d\theta^2+
\sin^2\theta d\varphi^2)\right),\eqno(5.1)
$$
where $\sigma=G$. The corresponding $\upsilon$-adic action is
$$
S_\upsilon[a]=\frac{1}{2}\int\limits_0\limits^1
dtNa^2(t)\left(-\frac{\dot a^2}{N^2a^2}+\frac{1}{a^2}-\lambda\right),
\eqno(5.2)
$$
where $\lambda=\Lambda\sigma^2$. The Euler-Lagrange equation of motion 
$$
\ddot a-N^2a\lambda=0
$$
has the solution
$$
a(t)=\frac{1}{2\sinh(N\sqrt\lambda)}
\left((a_2-a_1
e^{-N\sqrt\lambda})e^{N\sqrt\lambda t}+
(a_1e^{N\sqrt\lambda}-a_2)e^{-N\sqrt\lambda t}\right),
\eqno(5.3)
$$
where the boundary conditions are: $a(0)=a_1$, $a(1)=a_2$. For the
classical 
action it gives
$$
S_\upsilon^{cl}(a_2,N|a_1,0)=\frac{1}{2\sqrt\lambda}
\left[
N\sqrt\lambda+\lambda\left(\frac{2a_1a_2}{\sinh(N\sqrt\lambda)}-
\frac{a_1^2+a_2^2}{\tanh(N\sqrt\lambda)}\right)\right].\eqno(5.4)
$$
Quantum-mechanical propagator has the form
$$
{\cal K}_\upsilon(a_2,N|a_1,0)=\lambda_\upsilon
\left(
-\frac{\sqrt\lambda}{2\sinh(N\sqrt\lambda)}
\right)
\left|
\frac{\sqrt\lambda}{\sinh(N\sqrt\lambda)}
\right|_\upsilon^{1/2}
\chi_\upsilon
\left(
-S^{cl}_\upsilon(a_2,N|a_1,0)
\right).\eqno(5.5)
$$

The equation (4.9), in a more explicit form, reads
$$
\lambda_p
\left(
-\frac{\sqrt\lambda}{2\sinh(N\sqrt\lambda)}
\right)
\left|
\frac{\sqrt\lambda}{\sinh(N\sqrt\lambda)}
\right|_p^{1/2}
\chi_p
\left(
-\frac{N}{2}+\frac{\sqrt\lambda}{2\tanh(N\sqrt\lambda)}a_2^2
\right)
$$
$$
\times\!\!
\int\limits_{|a_1|_p\leq 1}\!\chi_p
\left(
\frac{\sqrt\lambda}{2\tanh(N\sqrt\lambda)}a_1^2
-\frac{\sqrt\lambda}{\sinh(N\sqrt\lambda)}a_2a_1
\right)da_1
=\Omega(|a_2|_p).\eqno(5.6)
$$
Using lower part of the Gauss integral  (2.9) for  $\nu=0$, we obtain
$$
\Omega(|a_2|_p)=\chi_p
\left(
-\frac{N}{2}+\frac{\sqrt\lambda}{2}\tanh(N\sqrt\lambda)a_2^2
\right)
\Omega(|a_2|_p)\eqno(5.7)
$$
with condition $|\frac{\sqrt\lambda}{2\tanh(N\sqrt\lambda)}|_p>1$,
i.e. $|N|_p<1$. For  $p\neq2$, l.h.s. is equal to $\Omega(|a_2|_p)$ if
$|\lambda|_p\leq1$ holds. Applying also the upper part of (2.9) to 
(5.6), we have
$$
\lambda_p
\left(
-\frac{\sqrt\lambda}{2\sinh(N\sqrt\lambda)}
\right)
\left|
\frac{\sqrt\lambda}{\sinh(N\sqrt\lambda)}
\right|_p^{1/2}
$$
$$
\times
\chi_p
\left(
-\frac{N}{2}+\frac{\sqrt\lambda a_2^2}{2\coth(N\sqrt\lambda)}
\right)
\Omega
\left(
\left|
\frac{\sqrt\lambda a_2}{\sinh(N\sqrt\lambda)}
\right|_p
\right)
=\Omega(|a_2|_p).\eqno(5.8)
$$
It becomes an equality if condition $|N|_p\leq1$ takes place. Finally,
if we take into account  convergence domain of hyperbolic functions,
i.e.
$|x|_p\leq\frac{1}{p}, p\neq2,$$ |x|_2\leq\frac{1}{2^2}$, we
obtain $p$-adic vacuum eigenstates
$$
\psi_p(a,N)=
\cases
\Omega(|a|_p), & |N|_p\leq1,\quad|\lambda|_p^{1/2}\leq1/p,\\
\Omega(|a|_2), & |N|_2\leq\frac{1}{4},\quad|\lambda|_2^{1/2}\leq1.  
\endcases
\eqno(5.9)
$$

In the form containing $\Omega$-function we also have eigenstates
$$
\psi_p(a,N)=
\cases 
\Omega(p^\nu|a|_p),
& |N|_p\leq p^{-2\nu},|\lambda|_p^{1/2}\leq p^{2\nu-1},\\
\Omega(2^\nu|a|_2),
& |N|_2\leq 2^{-2-2\nu},|\lambda|_2^{1/2}\leq 2^{2\nu},
\endcases
\eqno(5.10)
$$
where $\nu=1,2,\dots$.

\vskip 12pt
\noindent
\it 5.2. p-Adic and adelic de Sitter model
\vskip 2pt
\rm The de Sitter minisuperspace model in quantum cosmology is the
simplest,
nontrivial and exactly soluble model. This model is given by the 
Einstein-Hilbert action with cosmological term (4.2) without matter
fields, and by the Robertson-Walker metric
$$
ds^2=\sigma^2(-N^2(t)dt^2+a^2(t)d\Omega^2_3),\eqno(5.11)
$$
where $\sigma^2=\frac{2G}{3\pi}$ and
$a(t)$ is the scale factor. Instead of (5.11) we 
shall use
$$
ds^2=\sigma^2\left(-\frac{N^2(t)}{q(t)}dt^2+q(t)d\Omega^2_3)\right),\eqno(
5.12)
$$
which was considered in the real case in [17] and leads to quadratic
action.
The corresponding $\upsilon$-adic action
for this one-dimensional minisuperspace
model is
$$
S_\upsilon[q]=\frac{1}{2}\int\limits_{t_1}\limits^{t_2}
dtN\left(-\frac{\dot q^2}{4N^2}-\lambda q+1\right),\eqno(5.13)
$$
where $\lambda=\frac{\Lambda\sigma^2}{3}$. The definite $p$-adic
integrals
of the form (5.13) are considered in [18] by means of antiderivatives
of analytic functions, and they formally have the
same structure like those  in the real case.
The classical equation of motion ($N=1$)
$$
\ddot q=2\lambda
$$
with the boundary conditions $q(0)=q_1$ and $q(T)=q_2$, ($T=t_2-t_1$)
gives
$$
q(t)=\lambda t^2+(\frac{q_2-q_1}{T}-\lambda T)t+q_1 .\eqno(5.14)
$$
After substitution (5.14) into (5.13) and integration,
one obtains that the classical action is
$$
S_\upsilon^{cl}(q_2,T|q_1,0)=\frac{\lambda^2T^3}{24}-
[\lambda(q_1+q_2)-2]\frac{T}{4}-\frac{(q_2-q_1)^2}{8T}.\eqno(5.15)
$$
Since (5.15) is quadratic in $q_2$ and $q_1$, quantum-mechanical
propagator
has the form
$$
{\cal K}_\upsilon(q_2,T|q_1,0)=
\frac{\lambda_\upsilon(-8T)}{|4T|_\upsilon^{1/2}}
\chi_\upsilon(-S_\upsilon^{cl}(q_2,T|q_1,0)).\eqno(5.16)
$$
The equation (4.9) reads
$$
\frac{\lambda_p(-8T)}
{|4T|_p^{1/2}}
\chi_p
\left(
-\frac{\lambda^2T^3}{24}-\frac{T}{2}+\frac{\lambda q_2T}{4}+
\frac{q_2^2}{8T}
\right)
$$
$$
\times
\int\limits_{|q_1|_p\leq 1}\chi_p
\left(
\frac{q_1^2}{8T}+\left(
\frac{\lambda T}{4}-\frac{q_2}{4T}
\right)q_1
\right)dq_1
=\Omega(|q_2|_p).\eqno(5.17)
$$

For the eigenstates we have
$$
\psi_p(q,T)=
\cases
\Omega(|q|_p), & |T|_p\leq1,|\lambda|_p\leq1, p\neq 2,\\
\Omega(|q|_2), & |T|_2\leq\frac{1}{2},|\lambda|_2\leq1.  
\endcases
\eqno(5.18)
$$
and
$$
\psi_p(q,T)\!\!=\!\!
\cases
\Omega(p^\nu|q|_p),
 & |T|_p\leq p^{-2\nu},|\lambda|_p\leq p^{3\nu}, p\neq 2, \\
\Omega(2^\nu|q|_2),
& |T|_2\leq 2^{-2\nu},|\lambda|_2\leq 2^{1+3\nu},
\endcases
\eqno(5.19)
$$
where $\nu=1,2,\dots$.

\vskip 12pt
\noindent
\it 5.3 p-Adic and adelic model with a homogeneous scalar field
\vskip 10pt
\rm This model over the field of real numbers was considered in the paper
[19].
Metric is
$$
ds^2=\sigma^2\left(-N^2(t)\frac{dt^2}{a^2(t)}+a^2(t)d\Omega^2_3\right),
\eqno(5.20)
$$
and the corresponding real and $p$-adic action for the gravitational and
homogeneous scalar field is
$$
S_\upsilon[a,\phi]=\frac{1}{2}\int\limits_0\limits^1dtN
\left[-\frac{a^2\dot a^2}{N^2}+\frac{a^4\dot\phi^2}{N^2}-
a^2V(\phi)+1\right],\eqno(5.21)
$$
$
\phi=\left(\frac{4\pi G}{3}\right)^{1/2}\Phi.
$
The change of variables $x=a^2$ $\cosh(2\phi)$ and
$y=a^2\sinh(2\phi)$ in class of models  with the
scalar field potential 
$
V(\phi)=\alpha\cosh(2\phi)+\beta\sinh(2\phi),
$
where $\alpha$ and $\beta$ are arbitrary parameters, leads to the action
$$
S_\upsilon[x,y]=\frac{1}{2}\int\limits_0\limits^1dtN
\left[\frac{1}{4N^2}(-\dot x^2+\dot y^2)-\alpha x-\beta
y+1\right].\eqno(5.22)
$$
Varying this action with respect to $x$ and $y$ for the boundary
conditions
$x(0)=x_1,y(0)=y_1,x(1)=x_2,y(1)=y_2$, gives equations of motion with
the solutions
$$
x(t)=\alpha N^2t^2+(x_2-x_1-\alpha N^2)t+x_1,
$$
$$
y(t)=-\beta N^2t^2+(y_2-y_1+\beta N^2)t+y_1.\eqno(5.23)
$$ 

The classical action for the solutions (5.23) is given by
$$
S_\upsilon^{cl}(x_2,y_2,N|x_1,y_1,0)=
\frac{\alpha^2-\beta^2}{24}N^3+
\frac{1}{4}(2-\alpha(x_1+x_2)-
$$
$$
\beta(y_1+y_2))N
+\frac{-(x_2-x_1)^2+(y_2-y_1)^2}{8N},\eqno(5.24)
$$
and applying the formula (4.7) we get the corresponding propagator
$$
{\cal K}_\upsilon(x_2,y_2,N|x_1,y_1,0)=
\frac{1}{|4N|_\upsilon}
\chi_\upsilon(-S_\upsilon^{cl}(x_2,y_2,N|x_1,y_1,0)).\eqno(5.25)
$$

\noindent
Vacuum state for this two-dimensional minisuperspace model has the form
$\Omega(|x|_p)\Omega(|y|_p)$, and equation (4.9) reads
$$
\int\limits_{|x_1|_p\leq1}\enskip\int\limits_{|y_1|_p\leq1}
{\cal K}_p(x_2,y_2,N|x_1,y_1,0)dx_1dy_1=
\Omega(|x_2|_p)\Omega(|y_2|_p).\eqno(5.26)
$$

We obtain eigenstates
$$
\psi_p(x,y,N)=
\cases
\enskip
\Omega(|x|_p)\Omega(|y|_p), 
|N|_p\leq1,|\alpha|_p\leq1,|\beta|_p\leq1,p\neq 2,\\
\Omega(|x|_2)\Omega(|y|_2),  |N|_2\leq\frac{1}{2},
|\alpha|_2\leq2,
|\beta|_p\leq2.
\endcases
\eqno(5.27)
$$
For the states of the form $\Omega(p^\nu|x|_p)\Omega(p^\mu|y|_p)$
relevant
integral is
$$
\int\limits_{|x_1|_p\leq p^{-\nu}}\enskip\int\limits_{|y_1|_p\leq 
p^{-\mu}}
{\cal K}_p(x_2,y_2,N|x_1,y_1,0)dx_1dy_1=
\Omega(p^\nu|x_2|_p)\Omega(p^\mu|y_2|_p).\eqno(5.28)
$$
The corresponding $p$-adic eigenfunctions are:
$$
\psi_p(x,y,N)=
\cases
\Omega(p^\nu|x|_p)\Omega(p^\mu|y|_p),
\;|N|_p\leq p^{-2\nu},\;\;
|N|_p\leq p^{-2\mu},|\alpha|_p\leq p^{3\nu},
|\beta|_p\leq p^{3\mu}
\\
\Omega(2^\nu|x|_2)\Omega(2^\mu|y|_2),\;|N|_2\leq 2^{-2\nu},\;\;
|N|_2\leq 2^{-2\mu},
|\alpha|_2\leq 2^{3\nu},|\beta|_2\leq 2^{3\mu}
\endcases
\eqno(5.29)
$$
where $\nu,\mu=1,2,3,\dots$.

\vskip 12pt
\noindent
\it 5.4 p-Adic and adelic Bianchi I (k=0) model (with  three scaling
factors)
\vskip 10pt

\rm
We also apply the above formalism to the three-dimensional anisotropic
minisuperspace model, investigated in the real case in  [20]. 
The metric 
$$
ds^2=\sigma^2\left[-\frac{N^2(t)}{a^2(t)}dt^2 
+ a^2(t)dx^2 +b^2(t)dy^2+c^2(t)dz^2\right]\eqno(5.31)
$$
leads to the action
$$
S_\upsilon[a,b,c]=\frac{1}{2}\int^1_0dt
\left[-\frac{a}{N}(\dot a\dot bc +a\dot b\dot c+\dot ab\dot c)
-Nbc\lambda\right]\eqno(5.32)
$$
which by the substitution
$$
x=\frac{bc+a^2}{2},\quad y=\frac{bc-a^2}{2},\quad\dot z^2=a^2\dot b\dot
c
\eqno(5.33)
$$
gives classical action and propagator in the form
$$
S_\upsilon^{cl}(x_2,y_2,z_2,N|x_1,y_1,z_1,0)=
$$
$$
-\frac{1}{4N}\left[(x_2-x_1)^2-(y_2-y_1)^2+2(z_2-z_1)^2\right]
-\frac{\lambda N}{4}\left[(x_1+x_2)+(y_1+y_2)\right].\eqno(5.34)
$$
$$
{\cal K}_\upsilon(x_2,y_2,z_2,N|x_1,y_1,z_1,0)=
\frac{\lambda_\upsilon(-2N)}{\left|4^{1/3}N\right|_p^{3/2}}
\times
\chi_\upsilon\left(-S_\upsilon^{cl}(x_2,y_2,z_2,N|x_1,y_1,z_1,0)
\right).\eqno(5.35)
$$

By an analogous way to the previous models we get $p$-adic eigenstates
$$
\psi_p(x,y,z,N)=
\cases
\Omega(|x|_p)\Omega(|y|_p)\Omega(|z|_p),
& |N|_p\leq1,\quad|\lambda|_p\leq1,\\
\Omega(|x|_2)\Omega(|y|_2)\Omega(|z|_2),
& |N|_2\leq\frac{1}{2},
\quad|\lambda|_2\leq2.
\endcases
\eqno(5.36)
$$
and
$$
\psi_p(x,y,z,N)=
$$
$$
\cases
\Omega(p^{\nu_1}|x|_p)\Omega(p^{\nu_2}|y|_p)\Omega(p^{\nu_3}|z|_p),\;\;
\quad|N|_p\leq p^{-2\nu_i},|\lambda|_p\leq
p^{3\nu_1},p^{3\nu_2},\\
\Omega(2^{\nu_1}|x|_2)\Omega(2^{\nu_2}|y|_2)\Omega(2^{\nu_3}|z|_2),\;\;
\quad|N|_2\leq2^{-2\nu_{1,2}-1},2^{-2\nu_3-2},
|\lambda|_2\leq2^{3\nu_{1,2}+1}
\endcases
\eqno(5.37)
$$
for all $\nu_i \in Z$.

\vskip 12pt
\noindent
{\bf 6. Conclusion}
\vskip 12pt

It is shown that there exist $p$-adic and adelic counterparts
of all the above four minisuperspace cosmological models. Their
adelic eigenfunctions have the form (4.8). In particular, there is
a place eigenstate  of the form
$$
\Psi (q^1, ..., q^n) = \prod_{\alpha=1}^n \psi_\infty (q_\infty^\alpha)
\prod_p
\prod_{\alpha-1}^n \Omega(|q_p^\alpha|_p),
\eqno(6.1)
$$
where $\psi_\infty(q_\infty^\alpha)$ is the wave function of the universe
in the
standard quantum cosmology. To interprete $\Psi (q^1, ..., q^n)$ 
consider $|\Psi (q^1, ..., q^n)|_\infty^2 $ in rational points
$q^1, ..., q^n$, that is 
$$
|\Psi(q^1,...,q^n)|_\infty^2 = \prod_{\alpha=1}^n |\psi_\infty(q^\alpha)
|_\infty^2 \prod_p \prod_{\alpha=1}^n \Omega(|q^\alpha|_p).
\eqno(6.2)
$$
Due to the properties of $\Omega(|q^\alpha|_p)$, we have
$$
|\Psi(q^1,...,q^n)|_\infty^2 =
\cases
\prod_{\alpha=1}^n |\psi_\infty(q^\alpha)|_\infty^2 , \quad
q^1,...,q^n \in Z , \\
0,  \quad  \quad  q^\alpha \in Q\backslash Z .
\endcases
\eqno(6.3)
$$
According to the usual interpretation of a wave function,
from (6.3) follows discreteness of quantities $q^\alpha$
$(\alpha=1,...,n)$ at the natural scale $h=c=G=1$.
This kind of discreteness is a $p$-adic effect and depends
on adelic quantum state.

\vskip 12pt
\noindent
{\bf Acknowledgement}
\vskip 12pt

B.D. is supported in part by RFFI grants 990100866 and 990100166.
\vskip 12pt

\noindent
{\bf References}
\vskip 12pt

\item{1.}
    B. Dragovich and Lj. Ne\v si\'c, {\it Facta Univ.} {\bf 1}, 223
(1996).
\item{2.}
    V. S. Vladimirov, I. V. Volovich and E. I. Zelenov, ``p-Adic Analysis
and
    Mathematical Physics'', World Scientific, Singapore 1994.
\item{3.}
	 I. V. Volovich, ``Number theory as the ultimate physical 
       theory'', CERN preprint, CERN-TH. 4781/87 (July 1987).   
\item{4.}
	I. V. Volovich, {\it Class. Quantum Grav.} {\bf 4}, L83 (1987).
\item{5.}
	P. G. O. Freund and E. Witten, {\it Phys. Lett.}
        {\bf B 199}, 191 (1987).
\item{6.}
	V. S. Vladimirov, I. V. Volovich, {\it Dokl. Akad. Nauk SSSR}
       {\bf 302}, 320 (1988).
\item{7.}
	C. Alacoque, P. Ruelle, E. Thiran, D. Verstegen and J. Weyers,
        {\it Phys. Lett.} {\bf B 211}, 59 (1988).
\item{8.}
	V. S. Vladimirov and I. V. Volovich, 
        {\it Commun. Math. Phys.} {\bf 123}, 659 (1989).
\item{9.}
	B. Dragovich, ``On Adelic Quantum Mechanics'', in preparation.
\item{10.}
	G. S. Djordjevi\'c and B. Dragovich, ``p-Adic and adelic
     harmonic oscillator with time-dependent frequency'',
   to appear in {\it Teor. Mat. Fiz.}.
\item{11.}
        B. Dragovich, {\it Teor. Mat. Fiz.} {\bf 101}, 349 (1994);
        {\it Int. J. Mod. Phys.} {\bf A 10}, 2349 (1995).
\item{12.}
	Gel'fand I. M., Graev M. I. and Piatetskii-Shapiro I. I.
       ``Representation Theory and Automorphic Functions'', Nauka,
Moskva,
        1966.
\item{13.}
	D. Wiltshire, in {\it Proceedings of the 8th Physics Summer School},
        edited by B. Robson, N. Visvanathan and W. S. Woolcock (1996). 
\item{14.}
	B. Dragovich, ``Adelic Generalization of Wave Function of the
Universe'', {\it Publ. Obs. Astron. Belgrade}  {\bf 49}, 143 (1995):
"Adelic Wave  Function of the Universe" in {\it Proceedings of the Third
A. Friedmann Int. Seminar on Gravitation and Cosmology}, St. Petersburg,
1995.
\item{15.}
	G. Djordjevi\'c and B. Dragovich, {\it Mod. Phys. Lett} {\bf A 12},
1455
      1997.
\item{16.}
	J. J. Halliwel and R. C. Myers, {\it Phys. Rev.} {\bf D 40}, 4011
(1989).
\item{17.}
	J. J. Halliwel and J. Louko, {\it Phys. Rev.} {\bf D 39}, 2206 (1989).
\item{18.}
	I. Ya. Aref'eva, B. Dragovich, P. H. Frampton and
        I. V. Volovich, {\it Int. J. Mod. Phys.} {\bf A 6}, 4341 (1991).
\item{19.}
	L. J. Garay, J. J. Halliwel and G. A. Mena Marug\' an, 
        {\it Phys. Rev.} {\bf D 43}, 2572 (1991).
\item{20.}
	A. Ishikawa and H. Ueda, {\it Int. J. Mod. Phys.} {\bf D 2}, 249
(1993).

\end